\documentclass[twocolumn,prb]{revtex4}
\usepackage{subfigure}
\usepackage{color}
\usepackage{array}
\usepackage{amsmath}
\usepackage{amssymb}
\usepackage{wasysym}
\usepackage{bm}
\usepackage{graphicx}
\usepackage{epsfig}
\usepackage{verbatim}

\usepackage{ulem}

\renewcommand{\>}{\rangle}
\renewcommand{\(}{\left(}
\renewcommand{\)}{\right)}
\renewcommand{\[}{\left[}
\renewcommand{\]}{\right]}
\renewcommand{\v}[1]{\vec{#1}} 

\renewcommand{\d}{\partial}

\newcommand{\header}[1]{\vspace{4pt}\noindent {\bf #1 -- }} 

\begin{document}

\title{Quantum oscillations from generic surface Fermi arcs and bulk chiral modes in Weyl semimetals}

\author{Yi Zhang${}^{1}$, Daniel Bulmash${}^{1}$, Pavan Hosur${}^{1}$, Andrew C. Potter${}^{2}$, and Ashvin Vishwanath${}^{2}$}
\affiliation{${}^{1}$ Department of Physics, Stanford University,
Stanford, California 94305, USA} \affiliation{${}^{2}$ Department of
Physics, University of California, Berkeley, California 94720, USA.}
\date{\today}

\begin{abstract}
We re-examine the question of quantum oscillations from surface
Fermi arcs and chiral modes in Weyl semimetals. By introducing two
tools - semiclassical phase-space quantization and a numerical
implementation of a layered construction of Weyl semimetals - we
discover several important generalizations to previous conclusions
that were implicitly tailored to the special case of identical Fermi
arcs on top and bottom surfaces. We show that the phase-space
quantization picture fixes an ambiguity in the previously utilized
energy-time quantization approach and correctly reproduces the
numerically calculated quantum oscillations for generic Weyl
semimetals with distinctly curved Fermi arcs on the two surfaces.
Based on these methods, we identify a `magic' magnetic-field angle
where quantum oscillations become independent of sample thickness,
with striking experimental implications. We also analyze the
stability of these quantum oscillations to disorder, and show that
the high-field oscillations are expected to persist in samples whose
thickness parametrically exceeds the quantum mean free path.
\end{abstract}

\maketitle

Weyl semimetals are three-dimensional quantum materials
characterized by a band gap that closes at isolated points, Weyl
nodes, in the Brillouin zone. Weyl nodes serve as sources of
quantized monopole fluxes of $\pm 2\pi$ Berry curvature, whose sign
defines a chirality $\chi = \pm 1$ for each node, and hence serves
as an example of quantum topology in the absence of a band
gap\cite{Wan2011,Turner2013}. At a spatial surface, the bulk band
topology produces unusual Fermi-arc surface states, whose Fermi
``surface" consists of disjoint arc segments that pairwise connect
surface projections of opposite chirality Weyl
nodes\cite{Wan2011,PavanWeylFO,Turner2013,Haldane2014}, and have
been observed in photoemission
experiments\cite{Hasan2015Weyl,Lv2015Weyl} and band-structure
calculations\cite{wsmab} on crystalline materials. Moreover, in the
presence of a magnetic field, $\v{B}$, Weyl nodes exhibit chiral
Landau level (LL) modes\cite{ABJanomaly} with field-independent
dispersion $\varepsilon_{\chi,0}=\chi v_{\parallel}k_{\parallel}$,
where $v_\parallel$, $k_\parallel$ are respectively the velocity and
momentum along $\v{B}$.

Recently, it was shown\cite{DrewNC} that an applied magnetic field
perpendicular to the surface of a Weyl semimetal drives a novel kind
of cyclotron orbit in which electrons slide along a Fermi-arc on the
top surface from $\chi=+1$ towards $\chi=-1$ Weyl nodes, transfers
to the bulk chiral LL mode of the $\chi=-1$ node on which they
propagate to the bottom surface, traverse the bottom Fermi-arc and
return to the top surface via the mode with the opposite chirality.
Ordinary cyclotron orbits around closed Fermi surfaces of metals are
routinely studied via quantum oscillations, periodic-in-$1/B$
modulations in the density of states that appear in various
thermodynamic and transport properties, and help unveil the detailed
structure of the underlying Fermi surface. Ref.~\onlinecite{DrewNC}
showed that the quantized energy levels arising from these mixed
surface and bulk cyclotron orbits indeed exhibit periodic quantum
oscillations, whose phase exhibits a characteristic dependence on
sample thickness that distinguishes them from conventional cyclotron
orbits, and hence offering a direct probe of the topological
connection between surface Fermi arcs and bulk Weyl bands.
Experimental evidence for such quantum oscillations was recently
reported in the Dirac semimetal Cd$_3$As$_2$.\cite{Moll2015} In
addition, transport experiments were proposed based on the
distinctive electronic properties of these cyclotron
orbits\cite{Baum2015}.

The semiclassical quantization of these cyclotron orbits in
Ref.~\onlinecite{DrewNC} was carried out through ``energy-time"
quantization, by demanding that the product of the energy,
$\varepsilon$, of the electron and the semiclassical time of the
orbit, $t$, equals an integer multiple of $2\pi$. Noting that
$t=\(2L_z+2k_{0}\ell_B^{2}\)/v$, where $k_0$ is the $k$-space arc
length of the Fermi arcs on the top and bottom surfaces, $L_z$ is
the sample thickness, $v$ is the Fermi velocity, and
$\ell_B=1/\sqrt{eB}$ is the magnetic length, the energy-time
quantization condition states that the $n^\text{th}$ quantized level
crosses the chemical potential, $\tilde{\mu}$ at field $B=B_n$:
\begin{align}
\frac{1}{B_n}=\frac{e}{k_{0}}\left[\frac{\pi
v}{\tilde{\mu}}\left(n+\gamma\right)-L_{z}\right]\label{eq:previous}
\end{align}
which occur periodically in $1/B$ with period $f = \frac{\pi e
v}{k_0\tilde\mu}$ and a thickness dependent phase offset: $\phi(L_z)
=2\pi\gamma-\frac{2\tilde{\mu}L_z}{v}$.

However, this approach leaves open a basic question: what is the
overall zero of energy for $\tilde{\mu}$? This issue is experimental
pertinent, as it effects the frequency, $f$, of the quantum
oscillations. We will show that the nature choice of the energy of
the bulk Weyl node corresponds to the special case, implicitly
assumed in Ref.~\onlinecite{DrewNC}, where the Fermi arcs on the top
and bottom surfaces are identical. More generically, however, the
Fermi arcs may have different shapes, and the zero of energy need
not coincide with the Weyl node energy.

To generalize the results of Ref.~\onlinecite{DrewNC} to include the
generic case with arbitrarily curved Fermi arcs, we adopt an
alternative phase-space quantization perspective in which the
integral of momentum times spatial displacement is equal to
$\oint\vec{p}\cdot d\vec{r}=2\pi (n+\gamma)$ for integer $n$ and a
constant quantum offset $\gamma$. Comparison to the energy-time
quantization transparently identified the zero of energy as where
the surface arcs enclose zero k-space area using appropriate
extrapolation from the chemical potential to lower energy. This
method also predicts an additional thickness dependent correction to
Eq.~\ref{eq:previous}, which is difficult to obtain from the
energy-time quantization perspective.

We next construct a tight-binding model based on a layered
construction\cite{PavanWeylFO} of a Weyl semimetal, which enables
the numerical simulation of Weyl semimetals with generic surface
arcs. Using a recursive Greens function method, we numerically
simulate the field dependence of the density of states in a magnetic
field, and confirm the semiclassical predictions of the phase-space
quantization scheme.

Finally, we discuss the experimental consequences of our results.
First, we identify a special set of `magic' angles of the magnetic
field, for which the length-dependence of the phase of the quantum
oscillations drops out. We explain how this effect enables a smoking
gun signature of quantum oscillations from surface Fermi arcs in
recently measured thin-film devices with non-parallel
surfaces\cite{Moll2015}. Second, we examine the effects of
impurities, and find that these quantum oscillations are
surprisingly resilient to disorder for sufficiently strong fields.
In contrast to conventional quantum oscillations, which are obscured
by disorder unless the cyclotron orbit is smaller than the quantum
mean free path $\ell_Q$, we find that for strong fields, quantum
oscillations from surface Fermi arcs and bulk chiral modes can
persist in samples whose thickness substantially exceeds $\ell_Q$.

\header{Semiclassical phase-space quantization} We begin by
revisiting the semiclassical quantization of cyclotron orbits, which
generically demands that the phase difference between successive
Landau levels is equal to $2\pi$. The difference in phase
accumulated between two successive levels for a fixed magnetic field
can be expressed either in terms of the energy step and time or the
difference in the product of momentum and displacement:
\begin{align}
\Delta \phi=\int \Delta \varepsilon dt = \int \frac{\Delta
\varepsilon}{v_F}dr = \Delta \oint \v{p}\cdot d\v{r} = 2\pi
\label{etquantization}
\end{align}
where $v_F = \frac{\d \varepsilon}{\d p_\perp}$ is the Fermi
velocity, $p_\perp$ is the momentum perpendicular to the orbit, and
the last integral is over the spatial trajectory of the
semiclassical orbit. For a simple derivation via path integral, see
Appendix \ref{appendix:pathInt}. Importantly,
Eq.~\ref{etquantization} is expressed through the difference in
energy of neighboring Landau levels but makes no reference to their
absolute position. While the overall energy scale is unimportant
for, e.g. spectroscopy which probes only energy differences, quantum
oscillations experiments are conducted by varying $B$ at fixed
chemical potential $\mu$, such that the periodicity of quantum
oscillations depends explicitly on the ``zero of energy''.

Alternatively, the phase-space quantization framework offers an
unambiguous reference to the energy, in which the
momentum-displacement is integrated along the cyclotron orbit of
constant energy contour at a specific chemical potential. In
Appendix \ref{appendix:2deg}, we show how to reconcile these
methods, however, in the mean time we proceed with the more
transparent phase-space quantization approach.

\begin{figure}
\begin{centering}
\includegraphics[scale=0.40]{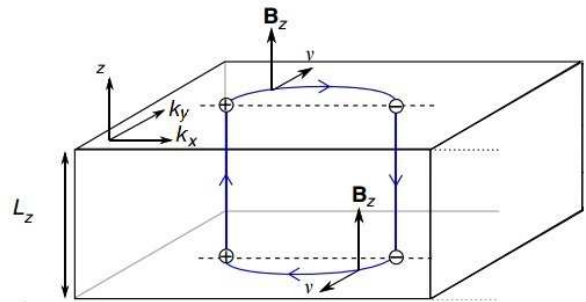}
\caption{Schematic plot of a semiclassical orbit of a Weyl semimetal
slab in a perpendicular magnetic field. The electrons traverse the
Fermi arc on the top surface, travel through the one-dimensional
chiral mode parallel to the magnetic field in the bulk, traverse the
corresponding Fermi arc on the bottom surface, and then return along
the opposite chiral mode through the bulk. Note that the real-space
orbit in the $x-y$ plane is rotated by $90^{\circ}$.} \label{fig5}
\end{centering}
\end{figure}

For the semiclassical cyclotron orbits described in
Ref.~\onlinecite{DrewNC}, the phase-space quantization condition is
$\oint\vec{p}\cdot d\vec{r}=\oint\vec{k}\cdot
d\vec{r}-e\oint\vec{A}\cdot d\vec{r}=2\pi\left(n+\gamma\right)$,
where the integral is over the four segments of the orbit: two Fermi
arcs on the surfaces and two chiral modes in the bulk parallel to
the magnetic field\footnote{There may exist additional phase
contributions at the turning points connecting the surface and bulk
orbits, which are presumably constant for large enough $L_z$ and can
be absorbed into the constant $\gamma$.}, as illustrated in Fig.
\ref{fig5}. Throughout, we choose the convention that chemical
potential, $\mu$, is measured from the energy of the Weyl nodes in
the bulk.

For the Fermi arcs, $\int\vec{p}\cdot d\vec{r}=e\Phi_z$, where
$\Phi_z$ is the magnetic flux contained within the real-space orbit
of area $S_R$ in the $x - y$ plane\cite{Onsager-Lifshitz}. The
semiclassical equations of motion imply
\begin{eqnarray}
\Phi_z & = & B_z S_{R}=B_z S_{k}\ell_B^{4}=\frac{S_{k}}{e^{2}B_z}
\end{eqnarray}
where $S_{k}$ is the $k$-space area enclosed by the two Fermi arcs
combined and $B_z$ is the $\hat z$ component of the magnetic field
$\vec B$. On the other hand, the chiral modes in the bulk are
parallel to the magnetic field, so $\int\vec{A}\cdot d\vec{r}=0$
and:
\begin{eqnarray}
\int\vec{p}\cdot d\vec{r}=\int\vec{k}\cdot d\vec{r}=
L_z\sec\theta\(\v{k}_W\cdot\hat{B}+2\frac{\mu}{v_\parallel}\)
\end{eqnarray}
where $\theta$ is the tilting angle of $\vec B$ from the surface
normal, $\v{k}_W$ is the wave vector from $+$ to $-$ chirality Weyl
nodes, and $\pm \frac{\mu}{v_\parallel}$ are the Fermi wave vectors
of their respective chiral modes with velocity $v_\parallel$
parallel to $\v{B}$, at chemical potential $\mu$. Adding the
contributions, phase-space quantization implies that quantum
oscillations occur at fields:
\begin{eqnarray}
\frac{1}{B_n} =
\frac{e}{S_k}\[2\pi(n+\gamma)\cos\theta-L_z\(\v{k}_W\cdot\hat{B}+\frac{2\mu}{v_\parallel}\)\]
\label{eq:qoresult}
\end{eqnarray}
Eq. \ref{eq:qoresult} is the main result of this paper.

When $\mu$ is close to the Weyl nodes and the Fermi velocity $v_{s}$
is approximately constant along the surface Fermi arcs, we can
expand $S_{k}=S_{k,0}+k_{0}^{T}\mu/ v_{s}$, where $S_{k,0}$ and
$k^{T}_{0}$ are the enclosed $k$-space area and total length of the
combination of the two Fermi arcs from both surfaces for $\mu$ at
the Weyl nodes. The frequency of the quantum oscillations
$f=1/\Delta(\frac{1}{B})$ is
\begin{eqnarray}
f=S_k/2\pi e&=&\left(S_{k,0}+k_{0}^{T}\mu/
v_{s}\right)/2\pi e \nonumber \\&=& k_{0}^{T}
\left(\mu_0+\mu\right)/2\pi e v_{s}\label{eq:qofreq}
\end{eqnarray}
where $\mu_{0}=S_{k,0}v_{s}/k_{0}^{T}$. We see that our results
reduce to those of Ref.~\onlinecite{DrewNC}, under the special
conditions: $S_k(\mu=0)=0$, $\v{k}_W\cdot\hat{B}=0$, and
$v_s=v_\parallel$. However, the phase-space quantization method
reveals two important generalizations:

(1) The $\tilde \mu$ defined in Eq. \ref{eq:previous} is generically
not measured from the energy of the bulk Weyl nodes. In particular,
if we require that $\mu$ is measured from the Weyl nodes, Eq.
\ref{eq:previous} should be modified by an offset
$\tilde\mu=\mu+\mu_{0}$. This reconciles the quantum oscillations
from phase-space quantization and energy-time quantization: the
contribution from the area $S_{k,0}$ enclosed by the Fermi arcs at
$\mu=0$ is reflected in $\mu_{0}$ while the contribution from the
area change $S_{k}-S_{k,0}=k_{0}^{T}\mu/v_{s}$ is reflected in
$\mu$. For cases where the area $S_{k,0}$ is large in comparison
with the area change, the inclusion of $\mu_{0}$ is necessary for
the correct interpretation of the quantum oscillations.

It is natural that $\mu_0$ should depend only on linearized
Fermi-surface properties such as the area enclosed and the Fermi
velocity, as the quantum oscillations generically encode only these
low-energy universal features. We note that since $v_s$ can in
principle depend on chemical potential, $\mu$, so does $\mu_0$ as
defined above. For a quadratic surface dispersion, $-\mu_0$ can be
interpreted as the energy (relative to the bulk Weyl nodes) at which
the surface arcs enclose zero area perpendicular to the magnetic
field. More generally, as we show in Appendix~\ref{appendix:2deg},
the appropriate way to reconcile energy-time quantization is to set
the zero of energy $-\mu_0$ at $\mu_0 = \frac{S_k}{\d S_k/\d\mu}$,
as the zero-area energy linearly extrapolated using the
Fermi-surface property $\frac{\d S_k}{\d\mu}\approx
\frac{k_0^T}{v_s}$.

(2) The thickness of the Weyl semimetal slab $L_z$ contributes to
the quantum oscillations through the phase offset of
$\phi(L_z)=\(\v{k}_W\cdot\hat{B}+\frac{2\mu}{v_\parallel}\)L_z\sec\theta$,
which shifts the $1/B$ positions of the quantum oscillation peaks.
Comparing to Ref.~\onlinecite{DrewNC}, we see that the thickness
dependent phase receives a contribution not only from the time
$t_\text{bulk}=\frac{L_z\sec\theta}{v_\parallel}$ taken to traverse
the bulk via the chiral mode, but also from the momentum-space
separation of the Weyl nodes projected onto $\vec{B}$.
Interestingly, for fixed chemical potential $\mu$, there exists a
special cone of angles of $\v{B}$, defined by:
$\v{k}_W\cdot\hat{B}=-\frac{2\mu}{v_\parallel}$, for which the phase
vanishes, $\phi(L_z)=0$, for all $L_z$, such that the oscillations
become independent of sample thickness.

\header{Numerical results and lattice model of Weyl semimetals} To
verify the semiclassical predictions from the phase-space
quantization approach we study a simple lattice model of Weyl
semimetal following the layered prescription in Ref.
\onlinecite{PavanWeylFO} and numerically calculate the density of
states $\rho(\mu,\v{B})$ in a slab geometry. In the absence of the
magnetic field, the Weyl semimetal is characterized by the following
Hamiltonian:
\begin{equation}
H_{\vec{k}}=\sum^{L_{z}}_{z=1}\left(-1\right)^{z-1}\varepsilon_{\vec{k}}c_{\vec{k},z}^{\dagger}c_{\vec{k},z}^{\vphantom\dagger}
+\sum^{L_{z}-1}_{z=1}h_{\vec{k},z}c_{\vec{k},z+1}^{\dagger}c_{\vec{k},z}^{\vphantom\dagger}+\mbox{H.c.}
\end{equation}
where $\vec{k} = \left(k_x, k_y\right)$ and the total number of
layers $L_{z}$ is odd so that the Fermi arcs on the top and bottom
surfaces can be different\cite{PavanWeylFO}. We consider an in-plane
dispersion $\varepsilon_{\vec{k}}=-2\cos k_{x}-2\cos
k_{y}+\varepsilon_{0}$ that represents nearest-neighbor hopping of
amplitude $-1$ and an on-site energy of $\varepsilon_{0}$.
$h_{\vec{k},z}$ represents nearest-neighbor interlayer hopping with
$h_{\vec{k},z}=-t\sin k_{y}-t_{0}$ if $z$ is odd and
$h_{\vec{k},z}=t\lambda\sin k_{y}+t_{0}$ if $z$ is even. We choose
$\lambda>1$, which ensures $h_{\vec{k},2}>h_{\vec{k},1}$ if
$k_{y}>0$ and vice versa. This model generates two Weyl nodes at
$(\pm k_{x}^{0},0,0)$ where $k_{x}^{0}$ is the in-plane Fermi wave
vector of $\varepsilon_{\vec{k}}$ along the $\hat{x}$ direction.
By definition $\vec{k}_W=2k_{x}^{0}\hat{x}$. The surface Fermi arcs
and the bulk chiral modes following the Weyl nodes are schematically
consistent with the geometry in Fig. \ref{fig5}.

In the presence of a magnetic field $\vec{B}$, the translation
symmetry in the $\hat y$ direction is preserved in the Landau gauge
$\vec{A}=\left(0, \Phi_z x - \Phi_x z, -\Phi_y x\right)$, where
$\Phi_{i}$ is the flux per plaquette perpendicular to the $\hat i$
direction in units of the magnetic flux quantum magnetic flux
quantum, $\Phi_0=\frac{h}{e}$, $i = x, y, z$. The Hamiltonian
becomes:
\begin{eqnarray}
H_{k_{y}}&=&\sum_{x,z}h_{\pi_y(x,z+1/2),z} \left(e^{-i\Phi_y
x}c_{x,z+1}^{\dagger}c_{x,z}^{\vphantom\dagger}+ \text{h.c.}\right)
\\
&-&\left(-1\right)^{z}\left[\left(\varepsilon_{0}-2\cos\pi_y(x,z)
\right)c_{x,z}^{\dagger}c_{x,z}^{\vphantom\dagger}-c_{x\pm1,z}^{\dagger}c_{x,z}^{\vphantom\dagger}\right]\nonumber
\label{eq:qoweylnumerics}
\end{eqnarray}
where $\pi_y(x,z) = k_y-A_y(x,z)$.

\begin{figure}
\begin{centering}
\includegraphics[scale=0.30]{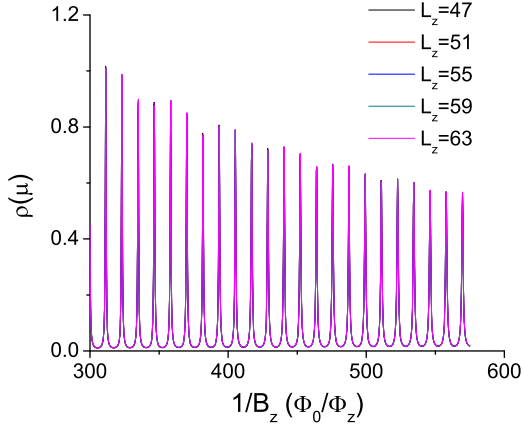} \caption{The density of
states $\rho(\mu)$ versus the inverse magnetic field $1/B_z$ (in
unit of $\Phi_0/\Phi_z$) for a Weyl semimetal slab of various
thickness $L_z$ shows clear quantum oscillations. The chemical
potential $\mu=0$ is at the Weyl nodes. The characteristic quantum
oscillation period is $\Delta(\Phi_0/\Phi_z)=11.74$.}\label{fig1}
\end{centering}
\end{figure}

The properties of this Hamiltonian such as the density of states
$\rho(\mu)=-\frac{1}{\pi L_x L_z}\sum_{x,z}\text{Im }
G(x,z;x,z;\mu)$ at the chemical potential $\mu$ can be calculated
with the recursive Green's function method where the real space
degrees of freedom in the $\hat{x}$ direction are treated
recursively\cite{recursiveGF, recursiveQO, harper}. For an
incommensurate flux $\Phi$, physical properties of $H_{k_{y}}$
between different choices of $k_{y}$ are equivalent in the
thermodynamic limit\cite{recursiveQO, harper} and the summation over
$k_{y}$ can be neglected.

We choose parameters $\varepsilon_{0}=3.0$, $t=1.0$, $t_{0}=2.0$,
$\lambda=2.0$, and a small imaginary part $\delta=0.001$ in addition
to the chemical potential $\mu$ as the level broadening. In this
model, although the chemical potential is at the Weyl nodes, the
Fermi arcs enclose a $k$-space area of $8.515\%$ of the surface
Brillouin zone. We first consider a magnetic field purely in the
$\hat z$ direction. Eq. \ref{eq:previous} would predict no quantum
oscillations if one assumed $\tilde \mu=0$, while Eq.
\ref{eq:qoresult} predicts quantum oscillations with a period
$\Delta(\Phi_0/\Phi_z) = 11.74$. The numerical results of the
density of states $\rho(\mu)$ versus the inverse magnetic field
$1/B$ and various slab thickness $L_z$, shown in Fig. \ref{fig1},
show clear signatures of quantum oscillations whose period is in
quantitative agreement with our formula.

\begin{figure}
\begin{centering}
\includegraphics[scale=0.30]{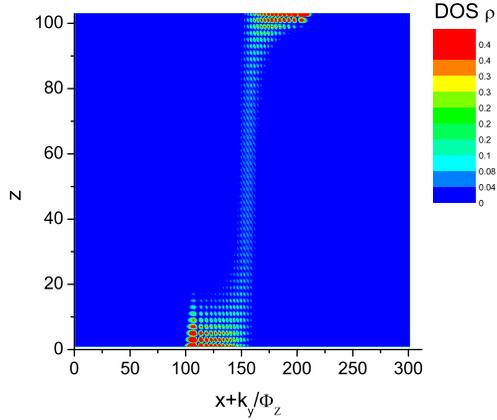} \caption{The local density of states
distribution in the $x-z$ plane at $\Phi_0/\Phi_z=311.40$, $\mu=0$
and $L_z=103$ is consistent with the cyclotron orbit illustrated in
Fig. \ref{fig5} and clearly consists of components from the Fermi
arcs on both Fermi surfaces and chiral LL modes in the
bulk.}\label{fig4}
\end{centering}
\end{figure}

To verify that the semiclassical orbit contains components in the
bulk as well as on both of the top and bottom surfaces, we calculate
the local density of states distribution
$\rho(x,z,\mu)=-\frac{1}{\pi} \text{Im } G(x,z;x,z;\mu)$ in the
$x-z$ plane with the $x$ coordinate replaced by $x+k_y/\Phi_z$. The
result for $\Phi_0/\Phi_z=311.40$, $\mu=0$ and $L_z=103$ is shown in
Fig. \ref{fig4}. The Fermi arcs are at $k_y>0$ and $k_y<0$ for the
top and bottom surfaces as well as the chiral modes in the bulk are
clearly visible.

\begin{figure}
\begin{centering}
\includegraphics[scale=0.30]{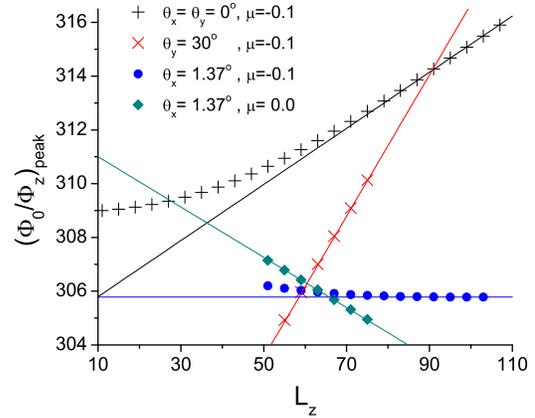}
\caption{Symbols mark the location of one of the density of states
peaks as a function of the slab thickness $L_z$ at different
chemical potential $\mu$ and magnetic field tilting angle $\theta_y$
or $\theta_x$. The lines are the asymptotic expression in the large
$L_z$ limit derived from the positions of and the Fermi velocity
around the Weyl nodes in the bulk:
$\left(\frac{\Phi_0}{\Phi_z}\right)_{\mbox{peak}}= \mbox{const.} +
L_z \cdot \frac{-\mu}{2\pi
t_0}\cdot\Delta\left(\frac{\Phi_0}{\Phi_z}\right)$ for a magnetic
field in the $\hat z$ direction, and refer to Appendix
\ref{appendix:tilttheory} for the expressions in the presence of a
tilted magnetic field. While the peak positions typically show
strong $L_z$ dependence, notably, for particular `magic' angles
(blue circles), the peak positions asymptotically become nearly
independent of sample thickness.}\label{fig3}
\end{centering}
\end{figure}

In addition, Eq. \ref{eq:qoresult} suggests that the thickness of
the slab $L_{z}$ changes the phase of the quantum oscillations, and
thus the actual locations of the $\rho(\mu)$ peaks in
$\Phi_0/\Phi_z$. For a magnetic field in the $\hat z$ direction and
$\mu=0$, $\v{k}_W\cdot\hat{B}=0$, we expect no $L_{z}$ dependence,
which is confirmed in Fig. \ref{fig1}. For a finite $\mu$ and a
field in the $\hat z$ direction, however, the shift $\delta(1/B)$ of
the peak positions is given by
\begin{equation}
\delta\left(\frac{\Phi_0}{\Phi_z}\right)=-\frac{\mu}{
v_{z}}\frac{2\delta
L_z}{2\pi}\cdot\Delta\left(\frac{\Phi_0}{\Phi_z}\right)=-\delta L_z
\cdot \frac{\mu}{2\pi
t_0}\cdot\Delta\left(\frac{\Phi_0}{\Phi_z}\right)\label{eq:lzdepend}
\end{equation}
where $\Delta\left(\frac{\Phi_0}{\Phi_z}\right)$ is the period of
the quantum oscillations. We numerically observe this shift in the
locations of the quantum oscillation peaks at $\mu =-0.1$ in Fig.
\ref{fig3}, where the location of one of the peaks is tracked as
$L_z$ is varied. The deviation from Eq. \ref{eq:lzdepend} at small
$L_z$ is due to the finite extent of the edge states (Fig.
\ref{fig4}). At relatively large $L_z$ where the physics in the
center of the slab can be approximately treated as in the bulk, Eq.
\ref{eq:lzdepend} gives an accurate description of the $L_z$
dependence of the quantum oscillation phenomena. The above
conclusions also hold true for a magnetic field that is tilted in
the $\hat y$ direction, e.g. $\vec B = B_z \left( \hat z + \hat y
\tan \theta_y\right)$, where only the $L_z$ coefficient is modified,
see Fig. \ref{fig3}.

In comparison, the magnetic field tilted in the $\hat x$ direction
gives qualitatively different behavior, since
$\v{k}_W\cdot\hat{B}\neq 0$ along $\vec B = B_z\left( \hat z + \hat
x \tan \theta_x\right)$. First, there exists $L_z$ dependence
$\delta\left(\frac{\Phi_0}{\Phi_z}\right)=-\delta L_{z}\cdot
k_{x}^{0}\tan\theta_x
\cdot\Delta\left(\frac{\Phi_0}{\Phi_z}\right)/\pi$ for chemical
potential $\mu=0$ at the energy of the Weyl nodes. Interestingly,
for a given chemical potential $\mu$, such $L_z$ dependence vanishes
at a special tilting angle $\theta_x^{(0)}$, which satisfies
\begin{equation}
\tan\theta_x^{(0)}=-\frac{\mu}{t_0}\left[4\left(k_{x}^{0}\right)^{2}-\mu^{2}/\sin^{2}\left(k_{x}^{0}\right)\right]^{-1/2}\label{eq:lztheta0}
\end{equation}

Numerical results for $\mu=0$ and $\mu=-0.1$ confirm these
expectations at tilting angle $\theta_x^{(0)}\approx 0.024$ as in
Eq. \ref{eq:lztheta0} for $\mu=-0.1$ (Fig.~\ref{fig3}). Derivations
and further discussion on the $L_z$ dependence of the peak positions
in a tilted magnetic field $\vec B$ are in Appendix
\ref{appendix:tilttheory}.

\header{Experimental consequences} The new features revealed by the
phase-space quantization treatment have several implications for
experiments. As an example, the angle and thickness dependence of
$\phi(L_z)$, observable by tracking individual quantum oscillation
peaks as a function of field orientation, can be used to
quantitatively determine the $k$-space separation of the Weyl nodes
from quantum oscillation measurements, which can be challenging to
accurately extract from other probes such as photoemission.

Moreover, the existence of a special set of angles for which the
quantum oscillations become independent of sample thickness enables
the following test: In recent experiments\cite{Moll2015}, Moll et
al. observed surface state oscillations in the Dirac semimetal
Cd$_3$As$_2$ in thin film devices with parallel surfaces, which were
absent in triangular devices with non-parallel top and bottom
surfaces. The absence of oscillations in the latter triangular
samples can be attributed to the destructive interference of orbits
with different $L_z$, due to the variation of device thickness along
the triangle. The above computations predict that this geometric
interference effect would be quenched for fields along the set of
angles for which $\phi(L_z)=0$, resulting in a reemergence of
quantum oscillations. In contrast to the negative signature of not
observing quantum oscillations in a triangular device, which could
potentially arise from other extrinsic effects, such an observation
would provide a clear positive signature of the non-local nature of
the Weyl orbits. We note that observing this effect requires the
field angle to be controlled to angular precision $\delta\theta
\lesssim 1/k_WL_z$, which may require thin devices.

As a final application, the phase-space quantization formulation
above naturally reveals the effect of bulk disorder on dephasing the
quantum oscillations associated with Weyl orbits (see
Appendix~\ref{app:Disorder} for a detailed discussion of disorder
effects). For conventional magnetic orbits, any scattering from
impurities strongly suppresses quantum oscillations, requiring $B
\gg k_F/\ell_Q$ where $\ell_Q$ is the quantum mean-free path
(distinct from the transport mean-free path, $\ell_{tr}$, which only
includes large-momentum transfer scattering) and $k_F$ is the Fermi
wave vector. As the surface-arc portion of the Weyl orbits is
locally identical to conventional cyclotron motion, observing
oscillations in the presence of disorder requires: $B\gg
k_0^T/\ell_Q$. Naively, phase coherence along the bulk part
similarly requires samples thinner than the quantum mean-free path,
$L_z\ll \ell_Q$, a potentially stringent condition since while
typical Weyl materials have large $\ell_{tr}$,  $\ell_Q$ is
typically much shorter\cite{OngNM2015}. However, the chiral nature
of the bulk orbit along with the spatially correlated nature of
disorder in low-density semimetals makes the bulk portion of the
orbit more resilient to disorder effects.

Namely, for an electron traveling along a bulk chiral LL, a random
potential $V(\v{r})$, produces a local shift in the wave vector:
$\delta k_\parallel(\v{r}) = -\tilde{V}(\v{r}_\perp,z)/v_\parallel$,
where $\tilde{V}$ is the matrix element of the disorder potential in
the chiral mode localized within $\approx \ell_B$ of transverse
position $\v{r}_\perp$ in the $xy$-plane. The total phase
accumulated in this fashion is given by: $\delta \phi = \int_0^{L_z}
dz \(\delta k_\parallel(\v{r_\perp},z)-\delta
k_\parallel(\v{r_\perp}+\v{d},z)\)$, where the first term represents
the random phase acquired traveling from bottom to top surface along
the $+$ chiral mode, and the second represents that of the return
journey on the counter-propagating chiral mode of the opposite Weyl
node. Between these two bulk legs of the orbit, the electron travels
a spatial distance $d\approx k_0^T\ell_B^2$ as it slides along the
top surface Fermi arc. The typical disorder for low-density Weyl
semimetals is poorly screened Coulomb impurities, which produce a
potential that is spatially correlated over characteristic length
scale $\xi\approx \sqrt{\frac{\ell_{tr}}{\ell_Q}}k_F^{-1}\gg
k_F^{-1}$. For low-field, $d\gg \xi$, the two bulk legs of the orbit
sample uncorrelated $V$, and dephasing indeed kills the quantum
oscillations for $L_z > \ell_B$. However, for higher fields $eB\gg
k_0^T/\xi$, $d\ll \xi$ and the top-to-bottom and bottom-to-top legs
accumulate nearly canceling random phases, which leads to the much
weaker requirement on sample thickness, $L_z <
\(\frac{\xi}{d}\)^2\ell_Q \gg \ell_Q $. For example, in
Cd$_3$As$_2$, we estimate that the high-field regime is obtained for
relatively low fields on the order of a few Tesla, in reasonable
agreement with the observed field-scale at which surface-state
oscillations onset in recent experiments\cite{Moll2015}.

\header{Discussion} In conclusion, we have compared quantum
oscillations with respect to the inverse magnetic field $1/B$ or
chemical potential $\mu$. The accurate definition of chemical
potential and its reference point is vital for correctly converting
between and reconciling different semi-classical quantization
perspectives. For the quantum oscillations from the surface Fermi
arcs and bulk chiral modes in Weyl semimetals, the general reference
point of $\mu$ does not necessarily coincide with the Weyl points in
the bulk. We derived the quantum oscillations using phase-space
quantization conditions and proposed essential generalizations to
previous conclusions and experimental consequences for generic Weyl
semimetals. We also verified our claims numerically following the
layered prescription.

We acknowledge insightful discussions with Itamar Kimchi, and Steven
A. Kivelson. YZ was supported by the NSF under Grant No.
DMR-1265593. DB was supported by the NSF under Grant No. DGE-114747.
PH was supported by the David and Lucile Packard Foundation. AP was
supported by the Gordon and Betty Moore Foundations EPiQS
Initiative through Grant GBMF4307. AV was supported by ARO MURI
program W911NF-12-1-0461.

\appendix

\section{Path integral derivation of energy-time quantization}
\label{appendix:pathInt}

Suppose that we have a quantum system with Hamiltonian $H$ and corresponding classical action $S$. Let the eigenstates of $H$ be $|E \rangle$. Then the propagator $U(x,x';t)$ obeys
\begin{equation}
U(x,x;T) = \sum_E |\langle x | E\rangle|^2e^{-iET}
\label{eq:fullPropagator}
\end{equation}
Computing the propagator in the path integral instead, we can make a semiclassical approximation and assume that the only paths which contribute are those paths $x_{cl}(t)$ which start at $x$ and return there in time $T$ according to the classical equations of motion. In this approximation, we get an alternative expression for the same function:
\begin{equation}
U(x,x;T) \approx \mathcal{N} \sum_{x_{cl}(t)}e^{iS[x_{cl}(t)]}
\end{equation}
where $\mathcal{N}$ is a prefactor irrelevant to us.

Now come the main assumptions. Inspired by the free particle in a magnetic field, for which these assumptions definitely hold, assume that there are no nonstationary closed classical paths except when $T= nT_0$ where $T_0$ is some classical period. Furthermore, we assume that each classical path at $T=nT_0$ is just $n$ loops of a single closed orbit $x_{\alpha}(t)$ which traverses the loop once in time $T_0$. In this case, absorbing any prefactors into $\mathcal{N}$,
\begin{align}
U(x,x;T) &\approx \mathcal{N}\sum_{\alpha}\sum_n e^{inS[x_{\alpha}(t)]}\delta(T-nT_0)\\
&= \mathcal{N}\sum_{\alpha}\int d\omega e^{-i\omega T}\sum_n e^{in(\omega T_0 + S[x_{\alpha}])}\\
&= \mathcal{N}\sum_{\alpha,m} \int d\omega e^{-i\omega T}\delta(\omega T_0 + S[x_{\alpha}]-2\pi m)
\end{align}
Comparing to the exact propagator Eq. \ref{eq:fullPropagator}, we see then that energies are labeled by $m$ and $\alpha$ and are given by
\begin{equation}
E_{m,\alpha} = \frac{2\pi m -S[x_{\alpha}]}{T_0}
\end{equation}
This implies the energy-time quantization condition in Eq.
\ref{etquantization}, but also tells us more. Here $S[x_{\alpha}]$
explicitly depends on the in-field Hamiltonian and thus the zero of
energy for the Landau levels, so $E_{m,\alpha}$ is automatically
defined relative to the same zero. In particular, if we assume $T_0
\sim 1/B$, then in deriving a condition like Eq. \ref{eq:previous}
the energy $\mu$ must be defined relative to the Landau level
spectrum, which need not coincide with the natural zero of energy
for the $B=0$ band structure.

\section{Quantum oscillations from energy-time
quantization}\label{appendix:2deg}

In this appendix, we compare the quantum oscillations from the two
different perspectives of energy-time quantization and phase-space
quantization. The phase-space quantization gives the condition of
the allowed semiclassical orbits:
\begin{equation}
\oint(\vec{k}-e\vec{A})\cdot\mathrm{d}\vec{r}=2\pi(n+\gamma)
\end{equation}
where the integral is over a constant energy contour in $k$-space at
certain chemical potential $\mu$. With the semiclassical equations
of motion, this can be re-written as:
\begin{equation}
\ell_{B}^2 S_k = \frac{\ell_{B}^2}{2}\ointop
k_{\perp}\mathrm{d}k_{\parallel}=2\pi(n+\gamma)\label{eq:pq-quant}
\end{equation}
where $k_\perp$ and $k_\parallel$ are the wave vectors normal and
parallel to the constant energy contour, respectively, $S_k$ is the
enclosed $k$-space area and $\ell_{B}=1/\sqrt{eB}$ is the magnetic
length.

On the other hand, consider a dispersion $\varepsilon(\vec{k})$ in
two dimensions, the time needed to complete a cyclotron orbit at
energy $\varepsilon_n$ is
\begin{equation}
t=\ell_{B}^2\oint\frac{\mathrm{d}k_{\parallel}}{v_\perp(\vec{k})}
\end{equation}
where $v_\perp$ is the Fermi velocity perpendicular to the contour.
The energy-time quantization states that
$\left(\varepsilon_{n}-\mu_0\right) t=2\pi(n+\gamma)$, which
suggests:
\begin{equation}
\ell_{B}^2\oint\frac{\varepsilon_{n}-\mu_0}{v_\perp(\vec{k})}\mathrm{d}k_{\parallel}
=2\pi(n+\gamma)\label{eq:et-quant}
\end{equation}
where $\mu_0$ is a constant offset and $\gamma$ is a Berry phase
contribution.

In particular, the two conditions in Eq. \ref{eq:pq-quant} and
\ref{eq:et-quant} match when
\begin{equation}
\frac{\varepsilon_{n}-\mu_0}{v_\perp} = \frac{k_\perp}{2}
\end{equation}
For a parabolic dispersion, e.g., a two-dimensional electron gas
$\varepsilon_{\vec k}=k^{2}/2m$, this is consistent with $\mu_0=0$.
Namely, the zero of energy is at the bottom of the band.

For a more generic dispersion, we assume the Fermi velocity
$v_\perp$ is independent of $\varepsilon_n$ in a small range around
the chemical potential $\varepsilon_F$, then it is straightforward
to take a derivative with respect to $n$ and get:
\begin{eqnarray}
\mu_0 \frac{\mathrm{d}}{\mathrm{d} n} \oint
\frac{\mathrm{d}k_{\parallel}}{v_\perp(\vec{k})} =
\frac{\mathrm{d}}{2\mathrm{d}
n}\oint\frac{\mathrm{d}k_{\parallel}}{v_\perp(\vec{k})} =
\frac{\mathrm{d} S_k}{\mathrm{d} n} \nonumber
\end{eqnarray}

\begin{eqnarray}
\mu_0  = S_k /\oint
\frac{\mathrm{d}k_{\parallel}}{v_\perp(\vec{k})}=S_k\cdot
\frac{\mathrm{d}\mu}{\mathrm{d} S_k}
\end{eqnarray}
Physically, given the Fermi surface area $S_k$ and its derivative
$\frac{\mathrm{d} S_k}{\mathrm{d}\mu}$ with respect to the chemical
potential $\mu$ near the Fermi level, the linear extrapolation to
lower energies gives the zero of energy as where the cross-section
area of the constant energy contour vanishes. For a linear
dispersion $\varepsilon=\pm v\left|\vec{k}\right|$ at chemical
potential $\varepsilon_F=v k_F$, for example, the zero of energy is
not at the Dirac node. It is straightforward to show that the
consistent quantum oscillations are derived from energy-time
quantization with $\mu_0 = \varepsilon_F / 2 = v k_F /2$.

Therefore, it is vital to understand where is the zero of energy
that the chemical potential $\mu$ is measured from. As an example of
the importance and ambiguity in correctly defining the zero of
energy, the quantum oscillations of a two-dimensional electron gas
$\varepsilon_{\vec k}=k^{2}/2m$ is:
\begin{equation}
\frac{1}{B_n}=\frac{2\pi
e}{S_{k}}\left(n+\frac{1}{2}\right)=\frac{2e}{k_{F}^{2}}\left(n+\frac{1}{2}\right)\label{eq:2degqo}
\end{equation}
where $S_{k}=\pi k_{F}^{2}$ and $k_{F}$ is the Fermi wave vector.

The energy-time quantization leads to the Landau levels:
\begin{eqnarray}
\varepsilon_{n} & = &
\omega_{c}\left(n+\gamma\right)-\mu_{0}\label{eq:2degll}
\end{eqnarray}
where $\omega_{c}=\frac{eB}{m}$ is the cyclotron frequency, and
$\mu_0$ and $\gamma$ are unknown constants since
$\Delta\varepsilon\times t=2\pi$ only gives the quantized level
spacings $\Delta\varepsilon=\varepsilon_{n}-\varepsilon_{n-1}$ and
contains no information on the exact zero of energy. For the quantum
oscillations at a fixed chemical potential $\mu=k_{F}^{2}/2m$, set
$\varepsilon_{n}=\mu$:
\begin{eqnarray}
\frac{k_{F}^{2}}{2m} & = & \frac{eB_n}{m}\left(n+\gamma\right)-\mu_{0}\nonumber \\
\frac{1}{B_n} & = &
\frac{2e}{k_{F}^{2}+2m\mu_{0}}\left(n+\gamma\right)
\end{eqnarray}
identical to Eq. \ref{eq:2degqo} if we set $\gamma=1/2$ and
$\mu_{0}=0$ as we have derived above. Importantly, the fundamental
behaviors of the quantum oscillations including its characteristic
frequency are not consistently recovered by this formula if
$\mu_{0}\neq 0$.

\section{Thickness dependence in a tilted magnetic field}\label{appendix:tilttheory}

We can linearize the dispersion near the Weyl nodes for the lattice
model we consider in the main text:
\begin{equation}
\varepsilon_{\vec{k}}^{\pm}=\left[\left(2t_{0}k_{z}\right)^{2}+\left(\left(\lambda-1\right)tk_{y}\right)^{2}+\left(\pm2\sin
k_{x}^{0}k_{x}\right)^{2}\right]^{1/2} \end{equation} where
$k_{x}^{0}=\cos^{-1}\left(\varepsilon_{0}/2-1\right)$ and the $\pm$
signs are for the two Weyl nodes with opposite chirality.

\begin{widetext}
For a magnetic field tilted in the $\hat{y}$ direction
$\vec{B}=B_{z}\left(\hat{z}+\hat{y}\tan\theta_y\right)$, the Fermi
wave vector of the chiral modes are
\begin{equation}
\vec{k}_{\parallel,1(2)}=\pm\frac{\mu}{2\left(\lambda-1\right)tt_{0}}\frac{4t_{0}^{2}\tan\theta_y\hat{y}+\left(\lambda-1\right)^{2}t^{2}\hat{z}}{\left[4t_{0}^{2}\tan^{2}\theta_y+\left(\lambda-1\right)^{2}t^{2}\right]^{1/2}}
\end{equation}
The shift of the peak positions is
\begin{align}
\delta\left(\frac{\Phi_0}{\Phi_z}\right) =-\frac{\mu \cdot \delta
L_{z}}{2\pi\left(\lambda-1\right)tt_{0}}
\left[4t_{0}^{2}\tan^{2}\theta_y+\left(\lambda-1\right)^{2}t^{2}\right]^{1/2}
\Delta\left(\frac{\Phi_0}{\Phi_z}\right)
\end{align}

Similarly, for a magnetic field tilted in the $\hat{x}$ direction
$\vec{B}=B_{z}\left(\hat{z}+\hat{x}\tan\theta_x\right)$, the Fermi
wave vector of the chiral modes are
\begin{equation}
\vec{k}_{\parallel,1(2)}=\pm\frac{\mu}{2t_{0}\sin
k_{x}^{0}}\frac{t_{0}^{2}\tan\theta_x\hat{x}+\sin^{2}k_{x}^{0}\hat{z}}{\left[t_{0}^{2}\tan^{2}\theta_x+\sin^{2}k_{x}^{0}t^{2}\right]^{1/2}}
\end{equation}
together with the location of the Weyl nodes at $\left(\pm
k_{x}^{0},0,0\right)$ therefore $\vec{k}_{W}=2k_{x}^{0}\hat{x}$, the
shift of the peak positions is
\begin{eqnarray}
\delta\left(\frac{\Phi_0}{\Phi_z}\right)&=&-\left[\frac{\mu}{2t_{0}\sin
k_{x}^{0}}\left(t_{0}^{2}\tan^{2}\theta_x+\sin^{2}k_{x}^{0}\right)^{1/2}+k_{x}^{0}\tan\theta_x\right]\nonumber
\frac{2\delta
L_{z}}{2\pi}\cdot\Delta\left(\frac{\Phi_0}{\Phi_z}\right)
\end{eqnarray}

\end{widetext}

There exists a residual $L_{z}$ dependence
$\delta\left(\frac{\Phi_0}{\Phi_z}\right)=-\delta L_{z}\cdot
k_{x}^{0}\tan\theta_x
\cdot\Delta\left(\frac{\Phi_0}{\Phi_z}\right)/\pi$ at $\mu=0$. In
addition, at tilting angle $\theta_x^{(0)}$ satisfying
\begin{equation}
\tan\theta_x^{(0)}=-\frac{\mu}{t_0}\left[4\left(k_{x}^{0}\right)^{2}-\mu^{2}/\sin^{2}\left(k_{x}^{0}\right)\right]^{-1/2}
\end{equation}
the coefficient in the square bracket vanishes, and the quantum
oscillations have no manifest $L_{z}$ dependence.

\section{Disorder Effects \label{app:Disorder}}
\subsection{Disorder Model}
As a model for the bulk effects of disorder, consider Weyl nodes with linear dispersion, and a random potential $V(r)$ with Gaussian distribution characterized by correlations $\overline{V(r)V(r')} = V_0^2f^{(3)}(r-r')$, where $f^{(d)}(r)$ is a smooth function that decays with characteristic length scale $\xi$, which, for concreteness, we will take as a normalized $d$-dimensional Gaussian with variance $\xi^2$, and $\overline{\(\cdots\)}$ indicates an average over disorder configurations.

In the Born approximation, the quantum lifetime, $\tau_Q$, characterizing the timescale between elastic scattering events in the bulk in the absence of a field is:
\begin{align}
\tau_Q^{-1} = \frac{2\pi\nu(0) V_0^2}{(k_F\xi)^2}
\end{align}
which is related to the quantum mean-free path by $\ell_Q = \frac{\tau_Q}{v}$, where $v_F$ is the geometric average of the different spatial components of the bulk velocity. Here $\nu(0) = \frac{k_F^2}{2\pi^2 v_F}$ is the density of states for the Fermi-surface of a single Weyl node, and $k_F = \frac{\mu}{v_F}$ is the Fermi wave vector for the Weyl pocket.

In contrast, the transport lifetime obtained by weighting scattering events between states with momenta $\v{k}$ and $\v{k}'$ by a factor of $\(1-\cos\theta_{\v{k},\v{k}'}\)\approx \(\frac{1}{k_F\xi}\)^2$, where $\theta_{\v{k},\v{k}'}$ is the angle between $\v{k},\v{k}'$, is given by:
\begin{align}
\tau_\text{tr}^{-1} = \frac{1}{(k_F\xi)^2}\tau_Q^{-1}
\end{align}
We can determine the parameter $V_0^2$ in terms of the measurable quantities $\tau_Q^{-1}$:
\begin{align}
V_0^2 = \frac{\pi v_F\xi^2}{\tau_Q}
\end{align}

\subsection{Different Scattering Processes}
For the bulk portion of the Weyl orbit, there are three potentially detrimental sources of disorder induced dephasing: 1) intervalley scattering between opposite chirality Weyl nodes, 2) scattering between different Landau levels (LLs) within a single Weyl node, and 3) random phase accumulated along the chiral LLs in the absence of inter-LL scattering. We will consider each of these channels in turn.
The rates for these dephasing channels add, indicating that their inverse length scales add: $\ell_\text{tot} = \(\sum_i \frac{1}{\ell_i}\)^{-1}$, indicating that the sample thickness limitation on observing quantum oscillations will be set by the shortest scattering length scale.

\subsection{Intervalley Scattering}
As intervalley scattering requires momentum transfer $\approx k_W$, for long-wavelength disorder, intervalley scattering will be suppressed by $\frac{1}{\(k_W\xi\)^2}$ compared to total quantum scattering, indicating:
\begin{align}
\ell_\text{inter-valley} \approx \(k_W\xi\)^2\ell_Q
\end{align}
In particular, since $k_W> k_F$, this length scale is even longer than the transport mean free path, $\ell_\text{tr}\approx (k_F\xi)^2\ell_Q$, for all field strengths.

\subsection{Inter Landau level Scattering}
For $\mu$ larger than the LL spacing, multiple non-chiral bulk LLs will coexist at the same energy as the chiral modes, and scattering between chiral and non-chiral modes within the same Weyl node is possible. However, since the wavefunctions of different LL modes differ on lengthscales of order $\ell_B$, for $\xi\gg \ell_B$, the matrix element for inter-LL scattering between levels with indices $m$ and $n$ is suppressed by approximately a factor of $\(\frac{\ell_B}{\xi}\)^{|m-n|}$:
\begin{align}
V_{n,m} &= \<u_n|V|u_m\>
\nonumber\\
&\approx \int dx \frac{e^{-x^2/2\ell_B^2}}{2\pi\ell_B^2} \sum_r H_n(x)H_m(x) \frac{x^rV^{(r)}(0)}{r!} \nonumber\\
&\approx \(\frac{\ell_B}{\xi}\)^{|m-n|}V_0
\end{align}
where $H_n(x)$ is the $n^\text{th}$ Hermite polynomial. The dominant mixing will hence come from minimal difference in LL indices, and hence, for $\ell_B\ll \xi$, the inter-LL scattering rate is suppressed by a factor of $\frac{|V_{0,1}|^2}{|V_0|^2}\approx \(\frac{\ell_B}{\xi}\)^2$, i.e. we expect:
\begin{align}
\ell_\text{inter-LL}\approx \(\frac{\xi}{\ell_B}\)^2\ell_Q
\end{align}
We will see below that this length scale is longer than that set by dephasing while propagating along the chiral bulk modes, which is expected to be the dominant limiting factor in observing quantum oscillations.

\subsection{Dephasing within the bulk chiral modes}
In the previous section, we have seen that inter-LL scattering may be neglected for $\ell_B\ll \xi$, and $L_z\ll \ell_\text{inter-LL}$. In this regime, the bulk portion of the orbit occurs purely within the chiral modes of the lowest Landau level. In the presence of an impurity potential $V(\v{r})$ that varies smoothly on the length scale of $k_F^{-1}$, we can model the $\pm$ chiral mode localized within $\ell_B$ of the guiding center position $\v{r}_\perp$ in the $xy$ plane, by the continuum Hamiltonian:
\begin{align}
H_{ch}^{(\pm)}(\v{r}_\perp) = \psi^\dagger \(\mp iv_F\d_z-U(\v{r}_\perp,z)\)\psi
\end{align}
where $U(\v{r}_\perp,z) = \int d^2\delta r_\perp V(r_0+\delta r_\perp,z)|u_0(r_\perp)|^2$ is the matrix element of $V$ within the lowest LL orbital with guiding center coordinate $\v{r}_\perp$ which has wave function $u_0(r_0+\delta r_\perp) \sim \frac{1}{4\pi \ell_B^2}e^{-\delta r_\perp^2/(4\ell_B^2)}$.

For moderate fields, $\ell_B \ll \xi$, the mean-square of these matrix elements is then given by:
\begin{align}
\overline{U(r,z)U(r,z')} & \approx  \frac{V_0^2}{2\pi\xi^2}f^{(1)}(z-z')
\end{align}
%
%

The random phase factor accumulated through the bulk portion of the orbit (ignoring mixing between chiral and non-chiral levels) is:
\begin{align}
e^{i\delta\phi}=\exp\[i\int_0^L \frac{U_{r_\perp}(z)-U_{r_\perp+d}(z)}{v_F}\]
\end{align}
where the first term comes from propagating from bottom to top surface along the $+$ chiral LL, and the second comes from returning from top to bottom surface along the counter-propagating $-$ chiral LL. In between the electron travels spatial distance $d=k_0^T\ell_B^2$ as it slides along the surface arc of the top surface.

Averaging the disorder phase over disorder gives:
\begin{align}
\overline{e^{i\delta\phi}} = e^{-\frac{1}{2v_F^2}\int_0^{L_z} dzdz' \overline{\(U(r,z)-U(r+d,z)\)\(U(r,z')-U(r+k_0\ell_B^2,z')\)}}
\end{align}
The suppression factor depends strongly on the ratio of the orbit size, $d=k_0^T\ell_B^2$, to the disorder correlation length.

\subsubsection{Low field regime $(d\gg \xi)$}
In the low field regime, $d\gg \xi$, $U(r,z)$ is uncorrelated with $U(r+k_0\ell_B^2,z)$, and:
\begin{align}
\overline{e^{i\delta\phi}} &\approx e^{-\frac{2}{v_F^2}\int_0^L dzdz' \overline{U(r,z)U(r,z')}}\approx e^{-\frac{v_F^2V_0^2}{\pi\xi^2} L_z}
\end{align}
from which we identify the relevant bulk ``mean-free path" length scale:
\begin{align}
\ell_{ch} = \frac{\pi v_F^2\xi^2}{V_0^2} = v_F\tau_Q = \ell_Q
\end{align}
which is just the quantum mean-free path.

\subsubsection{High field regime $(d\ll \xi)$}
On the other hand, for strong fields, $d\ll \xi$, the phase accumulated in traversing from bottom to top surfaces samples almost the same disorder configuration as the reverse trip, resulting in near cancellation of the total accumulated phase, and leading to a longer effective dephasing length for the chiral channel $\ell_{ch}\gg \ell_Q$.

To estimate $\ell_{ch}$, in this regime we also need the expression for:
\begin{align}
\overline{U(r,z)U(r+d,z')} \approx \[1-\(\frac{d}{\xi}\)^2\]\overline{U(r,z)U(r,z')}
\end{align}
with which we find:
\begin{align}
\overline{e^{i\delta\phi}} &\approx  e^{-\frac{1}{v_F^2}\int_0^{L_z} dzdz' \(\overline{U(r,z)U(r,z')-U(r,z)U(r+k_0\ell_B^2,z')}\)}
\nonumber\\
& \approx e^{-\(\frac{d}{\xi}\)^2\frac{L_z}{\ell_Q}} \equiv e^{-L_z/\ell_{ch}}
\end{align}

Hence, in the high-field regime, the chiral nature of the bulk LLs and long correlation length of disorder enables quantum oscillations to be observed for sample thicknesses up to
\begin{align}
\ell_{ch}\approx \(\frac{\xi}{d}\)^2\ell_Q
\end{align}
which can substantially exceed the quantum mean-free path.

We can express the enhancement of the dephasing length for the bulk chiral modes, $\ell_{ch}$,  compared to the quantum mean-free path, $\ell_Q$, in terms of measurable quantities including: 1) the ratio $\(\frac{\tau_{tr}}{\tau_{Q}}\)$ of transport to quantum lifetimes obtained respectively from transport and bulk quantum oscillation measurements, and 2) the frequency of the surface oscillations $f\approx k_Fk_0$, as:
\begin{align}
\frac{\ell_{ch}}{\ell_Q}=\(\frac{\xi}{d}\)^2 \approx \frac{\sqrt{\tau_\text{tr}/\tau_Q}}{f/B}
\end{align}

We note that the ``high-field" regime may be accessed for relatively
modest field scales. For example, for Cd$_3$As$_2$, $k_F\xi\approx
\sqrt{\frac{\tau_\text{tr}}{\tau_Q}}\approx 10-30$,~\cite{OngNM2015}
and $f\approx 60 T$,~\cite{Moll2015} and we estimate that $d\gg \xi$
can be achieved for fields of order a few Tesla, which is
incidentally consistent with the lowest fields for which surface
oscillations are seen in recent thin film devices\cite{Moll2015}.


\begin{thebibliography}{x}
\bibitem{Wan2011} Xiangang Wan, Ari M. Turner, Ashvin Vishwanath, and Sergey Y.
Savrasov, Phys. Rev. B 83, 205101 (2011).
\bibitem{Turner2013} Ari M. Turner, and Ashvin Vishwanath, arXiv-eprint:1301.0330 (2013).
\bibitem{Haldane2014} F.D.M. Haldane, arXiv-eprint:1401.0529 (2014).
\bibitem{PavanWeylFO} Pavan Hosur, Phys. Rev. B 86, 195102 (2012).
\bibitem{Hasan2015Weyl} Su-Yang Xu, et al., Science, 349, 613
(2015).
\bibitem{Lv2015Weyl} B. Q. Lv, et al., Phys. Rev. X 5, 031013
(2015).
\bibitem{wsmab} Hongming Weng, Chen Fang, Zhong Fang, B. Andrei Bernevig, and Xi
Dai, Physical Review X 5, 011029 (2015); Shin-Ming Huang, Su-Yang
Xu, et al., Nature Communications 6, 7373 (2015).
\bibitem{ABJanomaly} H.B. Nielsen, Masao Ninomiya, Physics Letters
130B, 6, 389 (1983).
\bibitem{DrewNC} Andrew C. Potter, Itamar Kimchi, Ashvin Vishwanath, Nature Communications 5, 5161(2014).
\bibitem{Moll2015} Philip J.W. Moll, Nityan L. Nair, Tony Helm, Andrew C. Potter, Itamar Kimchi, Ashvin Vishwanath, James G.
Analytis, arXiv-eprint:1505.02817 (2015).
\bibitem{Baum2015} Yuval Baum, Erez Berg, S. A. Parameswaran, Ady
Stern, arXiv-eprint:1508.03047 (2015).
\bibitem{Onsager-Lifshitz} L. Onsager, Phil. Mag. 43, 1006 (1952); I.M. Lifshitz and A.M. Kosevich, Sov. Phys. JETP 2, 636
(1956).
\bibitem{recursiveGF} Alexander Croy, Rudolf A. Roemer, Michael Schreiber, Lecture Notes in Computational Science and Engineerings, Springer, Berlin, 203(2006).
\bibitem{recursiveQO} Yi Zhang, Akash V. Maharaj, Steven A.
Kivelson, Physical Review B, 91, 085105 (2015).
\bibitem{harper} Yi Zhang, Danny Bulmash, Akash V. Maharaj, Chao-Ming Jian, and Steven
A. Kivelson, eprint-arXiv:1504.05205.
\bibitem{OngNM2015} Tian Liang, Quinn Gibson, Mazhar N. Ali, Minhao Liu, R. J. Cava, and N. P. Ong, Nature Materials 14, 280 (2015).
\end{thebibliography}
\end{document}